\documentclass[preprint,superscriptaddress,showpacs,preprintnumbers,amsmath,amssymb]{revtex4}

\usepackage{graphicx}
\usepackage{dcolumn}
\usepackage{bm}
\usepackage{color}

\begin{document}

\title{Solving the Dirac equation with nonlocal potential
by Imaginary Time Step method \footnote{Supported by the National
Key Basic Research Programme of China under Grant No 2007CB815000,
the National Natural Science Foundation of China under Grant No.
10775004.}}

\author{ZHANG Ying }
 \affiliation{State Key Lab Nucl. Phys. {\rm\&} Tech., School of Physics, Peking University, Beijing 100871, China}

\author{LIANG Hao-Zhao }
 \affiliation{State Key Lab Nucl. Phys. {\rm\&} Tech., School of Physics, Peking University, Beijing 100871, China}
 \affiliation{Institut de Physique Nucl\'eaire, IN2P3-CNRS and Universit\'e Paris-Sud,
    F-91406 Orsay Cedex, France}

\author{MENG Jie  \footnote{Email: mengj@pku.edu.cn}}
 \affiliation{State Key Lab Nucl. Phys. {\rm\&} Tech., School of Physics, Peking University, Beijing 100871, China}
 \affiliation{Department of Physics, University of Stellenbosch, Stellenbosch, South Africa}

\date{\today}

\begin{abstract}

The Imaginary Time Step (ITS) method is applied to solve the Dirac
equation with the nonlocal potential in coordinate space by the ITS
evolution for the corresponding Schr\"{o}dinger-like equation for
the upper component.  It is demonstrated that the ITS evolution can
be equivalently performed for the Schr\"{o}dinger-like equation with
or without localization.  The latter algorithm is recommended in the
application for the reason of simplicity and efficiency. The
feasibility and reliability of this algorithm are also illustrated
by taking the nucleus $^{16}$O as an example, where the same results
as the shooting method for the Dirac equation with localized
effective potentials are obtained.

\end{abstract}


\maketitle




With the operating of the worldwide new Radioactive Ion Beam (RIB)
facilities and the developments in the detection techniques, a lot
of novel aspects of nuclear structure and entirely unexpected
features have been found in the exploration of "exotic
nuclei"~\cite{Tanihata85,RIB01,Jonson04,Jensen04}. The extreme
neutron richness of exotic nuclei and the physics related to the low
density in the tails of nuclear matter distributions provide both
demand and challenge to solve complex many-body problem in
coordinate space.

As one of the best candidates for the description of exotic nuclei,
the Relativistic Mean Field (RMF) approach~\cite{Serot86} has
achieved lots of success in describing many nuclear phenomena during
the past years~\cite{Ring96,Vretenar05,MengPPNP}. In particular, the
Relativistic Continuum Hartree-Bogoliubov (RCHB)
theory~\cite{Meng98NPA} provides a fully self-consistent treatment
of pairing correlations in the presence of the continuum and thus
gives a reliable description of nuclei far away from the line of
$\beta$-stability, which suggests a new mechanism for the formation
of the halo phenomena~\cite{MengHaloPRL}.

Within the RMF approach, however, the Fock terms of the energy
density functional are neglected under the Hartree approximation.
The recent development of density-dependent relativistic
Hartree-Fock (DDRHF) theory~\cite{DDRHFPLB} resumes the Fock terms
and obtains quantitatively comparable precision with the RMF theory
for nuclear matter and finite nuclei~\cite{DDRHFPLB,DDRHFPKA}.
Moreover, within the DDRHF theory, the importance of the Fock terms
is evidenced by the improvement on the descriptions of the nuclear
shell structures~\cite{DDRHFPKA} and their
evolutions~\cite{DDRHFpi}, the excellent reproduction of the
spin-isospin resonance based on the self-consistent random phase
approximation~\cite{RPAPRL}, as well as the influences on isospin
properties of nuclear matter and neutron stars at high
densities~\cite{DDRHFsun}.  More recently, as an extension of the
DDRHF theory, the relativistic Hartree-Fock-Bogoliubov theory with
density dependent meson-nucleon couplings (DDRHFB) is developed for
the description of the exotic nuclei~\cite{DDRHFB}.

For the future study, one of the natural extensions of the RCHB or
DDRHFB theory is the exploration of deformed exotic nuclei, which
will provide the decisive conclusion on the existence of deformed
halo. Efforts along this line have been made in the past several
years. Due to the difficulty in solving the coupled channel
differential equations for deformed system in coordinate
space~\cite{Price87}, the method for expansion in Woods-Saxon basis
was proposed~\cite{ZhouWS03}, which becomes very time consuming for
the heavy system.

The Imaginary Time Step (ITS) method~\cite{ev8-80} is another
effective approach to solve the nonrelativistic problems in
coordinate space and has achieved lots of success in the
conventional mean field approach~\cite{ev8-cpc05}. More importantly,
its extension from spherical to deformed systems is straightforward.
It is therefore worthwhile to adopt this method for the relativistic
system. Naively, one may assume that the ITS method will result in a
great disaster due to the Dirac sea. Fortunately, it has been
demonstrated that this disaster can be avoided by the ITS evolution
for the Schr\"{o}dinger-like or charge-conjugate
Schr\"{o}dinger-like equation for the solution respectively in the
Fermi or Dirac sea~\cite{ITSDirac}.

For the application of the ITS method in the relativistic
Hartree-Fock theory, one has to solve the Dirac equation with
nonlocal potential due to the Fock terms. The corresponding Dirac
equation will be a set of coupled integro-differential equations,
which cannot be solved by the conventional shooting method directly.

In this letter, the ITS method will be applied to solve the Dirac
equation with the nonlocal potential in coordinate space following
the success in solving the Dirac equation with the local potential
by the ITS evolution for the corresponding Schr\"{o}dinger-like
equation for the upper component~\cite{ITSDirac}. It will be
demonstrated that the ITS evolution can be equivalently performed
for the Schr\"{o}dinger-like equation with or without localization,
and the latter algorithm is recommended for the reason of simplicity
and efficiency.


In the relativistic Hartree-Fock approach, the main issue is to
solve the Dirac equation
 \begin{equation}
    \left\{ \mbox{\boldmath$\alpha$}\cdot\mbox{\boldmath$p$}+\beta \left[ M+S(\mbox{\boldmath$r$})\right]+V(\mbox{\boldmath$r$})\right\} \varphi_a(\mbox{\boldmath$r$})
      +\int d\mbox{\boldmath$r$}' U(\mbox{\boldmath$r$}, \mbox{\boldmath$r$}') \varphi_a(\mbox{\boldmath$r$}') = \varepsilon_a\varphi_a(\mbox{\boldmath$r$}),
 \end{equation}
with the local (scalar and vector) potentials $S(r)$ and $V (r)$,
and the nonlocal potential $U(\mbox{\boldmath$r$},
\mbox{\boldmath$r$}')$. For clarity and simplicity, $V (r)\pm S(r)$
are assumed to be the spherical Woods-Saxon potentials similar as in
Ref. \cite{ZhouWS03}, and thus the Dirac spinor takes the form,

  \begin{equation}
   \varphi_a(\mbox{\boldmath$r$})= \dfrac{1}{r}\left( \begin{array}{c}
                    {iF_a(r)}\mathcal{Y}^{l_a}_{j_am_{j_a}}(\Omega)\\
                    -{G_a(r)}\mathcal{Y}^{l'_a}_{j_am_{j_a}}(\Omega)
                  \end{array}
           \right)\chi_{\frac{1}{2}}(q_a), ~ l_a + l'_a = 2j_a,
  \end{equation}
with the isospinor $\chi_{\frac{1}{2}}(q_a)$ and the spherical
spinor $\mathcal{Y}^l_{jm_j}(\Omega)$. The single-particle state is
labeled by the quantum numbers $\{n_a,l_a,\kappa_a,j_a\}$ of the
upper component, where $\kappa_a=\pm|j_a+1/2|$ for $l_a=j_a \pm
1/2$.

Similar as in Ref.~\cite{NonLocalJCP5}, the nonlocal potential
$U(\mbox{\boldmath$r$}, \mbox{\boldmath$r$}')$ is expanded as,

  \begin{equation}
   U(\mbox{\boldmath$r$}, \mbox{\boldmath$r$}') = \sum_l \dfrac{u_l(r,r')}{rr'}
  \dfrac{2l+1}{4\pi}P_l(\cos\omega), \label{Urrpexpand}
  \end{equation}
where $P_l$ is the Legendre polynomial and $\omega$ is the angle
between $\mbox{\boldmath$r$}$ and $\mbox{\boldmath$r$}'$. More
explicitly, if $U(\mbox{\boldmath$r$},\mbox{\boldmath$r$}')$ takes
the same form as in Ref.~\cite{NonLocalJCP5},
 \begin{equation}
  U(\mbox{\boldmath$r$},\mbox{\boldmath$r$}')= \dfrac{\sqrt{u(r)u(r')}}{\left(
  \pi\gamma^2\right)^{\frac{3}{2}}}e^{-{\left(\mbox{\boldmath$r$}-\mbox{\boldmath$r$}'\right)^2}/{\gamma^2}},
 \end{equation}
the expansion coefficient $u_l(r,r')$ can be obtained as,
 \begin{equation}
  u_l(r,r')
  = \dfrac{2\sqrt{rr'}}{\gamma^2}\sqrt{u(r)u(r')}e^{-\frac{r^2+r'^2}{\gamma^2}}
  I_{l+\frac{1}{2}}\left( \dfrac{2rr'}{\gamma^2} \right) \label{WS-besI},
 \end{equation}
 where $I_{l+\frac{1}{2}}$ is the modified Bessel function of the first
 kind. In practice, the central part $u(r)$ could take the shape of Woods-Saxon type,
 \begin{equation}
   u(r) = -\dfrac{V_0}{1+e^{(r-R)/\alpha}}, ~~ R=r_0A^{1/3},
   \label{WS}
 \end{equation}
 with the parameters, $V_0$, $r_0$, $\alpha$ and the nonlocality parameter $\gamma$.

With the expansion as in Equation (\ref{Urrpexpand}), the Dirac
equation with nonlocal potential can be reduced as
 \begin{equation}
  \left(\begin{array}{cc}
         V+S+M   & -\frac{d}{dr}+\frac{\kappa_a}{r} \\
         +\frac{d}{dr}+\frac{\kappa_a}{r} & V-(S+M)
        \end{array}\right)
  \left(\begin{array}{c}
         F_a(r) \\
         G_a(r)
        \end{array}\right) +
      \left(
         \begin{array}{c}
           X_a(r) \\
           Y_a(r)  \\
         \end{array}
       \right)
  =\varepsilon_a \left(\begin{array}{c}
         F_a(r) \\
         G_a(r)
        \end{array}\right), \label{dirac-2}
 \end{equation}
where the nonlocal terms are
\begin{subequations}
 \begin{eqnarray}
   X_a(r) &=& \int dr' U_{X}(r,r') F_a(r'), ~~U_{X}(r,r')\equiv u_{l_a}(r,r'), \label{NLXr} \\
   Y_a(r) &=& \int dr' U_{Y}(r,r') G_a(r'), ~~U_{Y}(r,r')\equiv u_{l'_a}(r,r').\label{NLYr}
 \end{eqnarray}
\end{subequations}

The coupled integro-differential equations in
Equation(\ref{dirac-2}) cannot be directly solved by the
conventional shooting (Runge-Kutta) method unless they are
transformed to homogeneous differential equations by localization
\cite{LDiracBouyssy}. For example, one can rewrite Equation
(\ref{dirac-2}) as
 \begin{equation}
  \left(\begin{array}{cc}
         V+S+M+X_{a,F_a}  & -\frac{d}{dr}+\frac{\kappa_a}{r}+X_{a,G_a} \\
         +\frac{d}{dr}+\frac{\kappa_a}{r}+Y_{a,F_a} & V-(S+M)+Y_{a,G_a}
        \end{array}\right)
  \left(\begin{array}{c}
         F_a(r) \\
         G_a(r)
        \end{array}\right)
  =\varepsilon_a \left(\begin{array}{c}
         F_a(r) \\
         G_a(r)
        \end{array}\right), \label{dirac-3}
 \end{equation}
with the localized effective potentials,
\begin{subequations}
 \begin{eqnarray}
   X_{a,F_a}  \equiv  \dfrac{F_a(r)X_a}{F^2_a(r)+G^2_a(r)},
   & &
   X_{a,G_a}  \equiv  \dfrac{G_a(r)X_a}{F^2_a(r)+G^2_a(r)},\\
   Y_{a,F_a}  \equiv  \dfrac{F_a(r)Y_a}{F^2_a(r)+G^2_a(r)},
   & &
   Y_{a,G_a}  \equiv  \dfrac{G_a(r)Y_a}{F^2_a(r)+G^2_a(r)},
 \end{eqnarray}
\end{subequations}
then Equation(\ref{dirac-3}) could be solved iteratively from an
initial guess of the wave function by the shooting method.

For the ITS method, as demonstrated in Ref.~\cite{ITSDirac}, the
evolution can be performed by using the effective single-particle
Hamiltonian for the upper component $\hat{h}_{\mbox{\tiny eff.}}$.
With the relation between the upper and lower components in
Equation~(\ref{dirac-3}),
 \begin{equation}
  G_a = \frac{1}{M_L}\left( \frac{dF_a}{dr}+\frac{\kappa_a}{r}F_a +Y_{a,F_a}F_a\right),
  \mbox{ where } ~ M_L = M-(V-S)-Y_{a,G_a}+\varepsilon_a, \label{SchLocalG}
 \end{equation}
the Schr\"{o}dinger-like equation for the upper components is,
$\hat{h}_{\mbox{\tiny eff.}}F_a=\varepsilon_a F_a$, with
 \begin{eqnarray}
  \hat{h}_{\mbox{\tiny eff.}}F_a &=&- \dfrac{1}{M_L}\dfrac{d^2F_a}{dr^2}+\left[
   -\dfrac{1}{M_L}(Y_{a,F_a}-X_{a,G_a})+\dfrac{1}{M_L^2}\dfrac{dM_L}{dr}\right]
   \dfrac{dF_a}{dr} \nonumber \\
   &+& \left[ \dfrac{1}{M_L}\dfrac{\kappa_a(\kappa_a+1)}{r^2}+\dfrac{1}{M_L}\dfrac{\kappa_a}{r}
   \left( \dfrac{1}{M_L}\dfrac{dM_L}{dr}+Y_{a,F_a}+X_{a,G_a}\right) \right. \nonumber \\
   &+& \left. \left( -\dfrac{1}{M_L}\dfrac{dY_{a,F_a}}{dr}+\dfrac{1}{M_L^2}\dfrac{dM_L}{dr}Y_{a,F_a}
   +\dfrac{1}{M_L}X_{a,G_a}Y_{a,F_a}\right) + (V+S+M+X_{a,F_a})\right] F_a.  \nonumber \\ \label{heffFL}
 \end{eqnarray}

On the other hand, one could also obtain the relation between the
upper and lower components from Equation~(\ref{dirac-2}),
   \begin{equation}
     G_a = \dfrac{1}{M_+}\left(
     \dfrac{dF_a}{dr}+\dfrac{\kappa_a}{r}F_a+Y_a\right), ~~\mbox{where}~~
     M_+=M-(V-S)+\varepsilon_a, \label{FtoG}
   \end{equation}
and the corresponding Schr\"{o}dinger-like equation for the upper
component gives
  \begin{eqnarray}
   \hat{h}_{\mbox{\tiny eff.}}  F_a&=&
   -\dfrac{1}{M_+}\dfrac{d^2F_a}{dr^2}+\dfrac{1}{M_+^2}\dfrac{dM_+}{dr}\dfrac{dF_a}{dr}
   +\left[ (V+S+M)+\dfrac{1}{M_+^2}\dfrac{dM_+}{dr}\dfrac{\kappa_a}{r}+
   \dfrac{1}{M_+}\dfrac{\kappa_a(\kappa_a+1)}{r^2}\right] F_a \nonumber \\
   & & -\dfrac{1}{M_+}\dfrac{dY_a}{dr}+\dfrac{1}{M_+^2}\dfrac{dM_+}{dr}Y_a
   +\dfrac{1}{M_+}\dfrac{\kappa_a}{r}Y_a+X_a.
   \label{heffFNL}
  \end{eqnarray}

Although the right-hand sides are different,
Equation~(\ref{heffFNL}) is exactly the same as
Equation~(\ref{heffFL}). Moreover, one should note that the ITS
evolution involves only the operation $\hat{h}_{\mbox{\tiny
eff.}}F_a$. Therefore, the nonlocalized ITS evolution for
Equation~(\ref{heffFNL}) is equivalent to the localized one for
Equation~(\ref{heffFL}). For the reason of simplicity and
efficiency, the nonlocalized ITS evolution for
Equation~(\ref{heffFNL}) will be used.


In the numerical calculations, the ITS evolution starts from the
initial orthogonal wave functions \{$\varphi_a^{(0)}$\} with
spherical Bessel function for the upper components and zero for the
lower components. Together with the given nonlocal potentials
(\ref{WS-besI}), the initial $X^{(0)}_a(r)$ and $Y^{(0)}_a(r)$ are
constructed according to Eqs. (\ref{NLXr}) and (\ref{NLYr}). Then
the operation $\hat{h}_{\mbox{\tiny eff.}}F_a$ in
Equation~(\ref{heffFNL}) is obtained and applied to evolve the
single-particle wave functions, which is carried out in coordinate
space within a spherical box $[0,R]$ with the box size $R=20$ fm,
mesh size $dr=0.1$~fm and a typical time step $\Delta t=10^{-26}$~s.
The set of wave functions thus obtained is then orthogonalized by
the standard Gram-Schmidt procedure to provide a new set of
orthogonal single-particle wave functions which will be used for the
next ITS evolution. This process is repeated till the final
convergence.

\begin{figure}[!h]
       \includegraphics[width=1.0\textwidth]{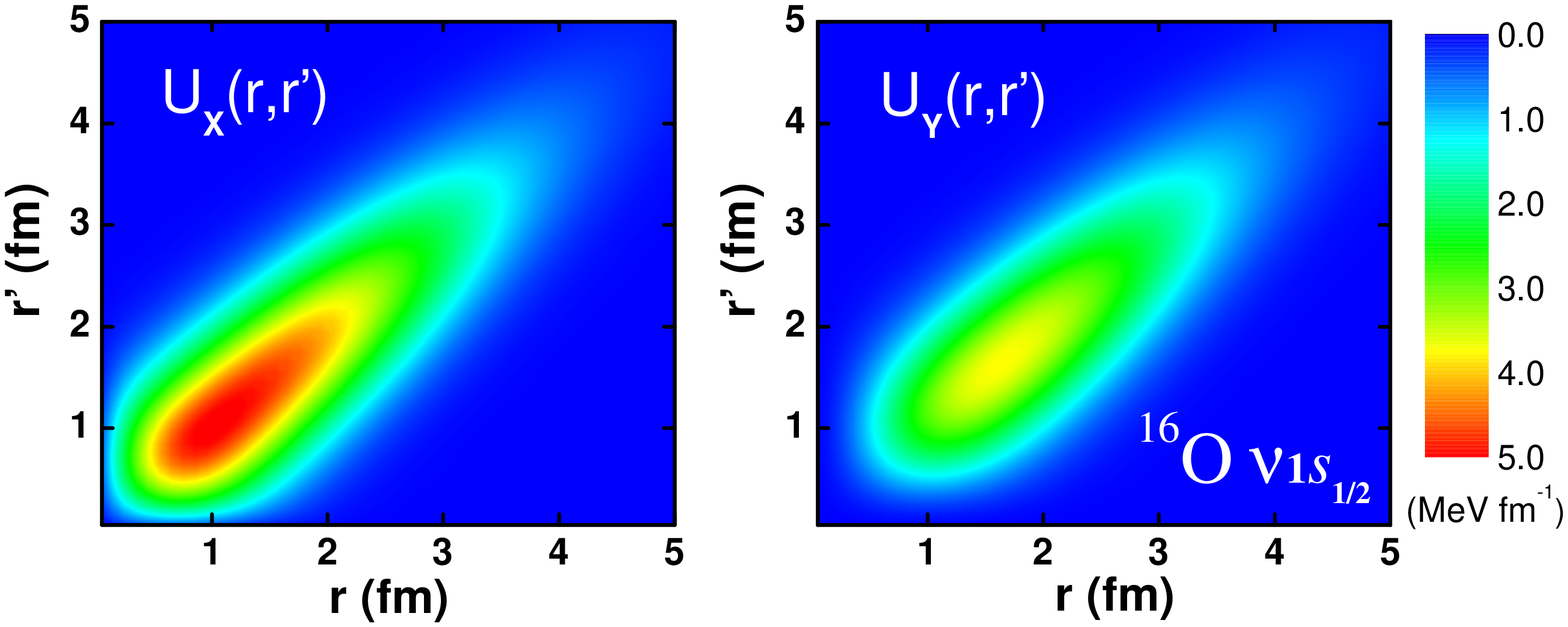} 
       \caption{ (color online)
                The nonlocal potentials $U_X(r,r')$ and $U_Y(r,r')$ for
                $\nu 1s_{1/2}$ state in nucleus $^{16}$O.
                The parameters in Eqs. (\ref{WS-besI}) and (\ref{WS})
                are chosen as $V_0=10$ MeV, $r_0=1.04$ fm, $\alpha=0.65$ fm and $\gamma=1.0$ fm.
       }\label{Fig1}
      \end{figure}

Take the state $\nu1 s_{1/2}$ in $^{16}$O as an example, the
nonlocal potentials $U_X(r,r')$ and $U_Y(r,r')$ defined in Eqs.
(\ref{NLXr}) and (\ref{NLYr}) are shown in Fig. \ref{Fig1}. The
detailed parameters for the nonlocal potentials chosen tentatively
to give a reasonable single-particle spectrum in $^{16}$O can be
found in the corresponding caption. The nonlocal potentials
$U_X(r,r')$ and $U_Y(r,r')$ are state-dependent with different
orbital angular momentum quantum numbers $l_a$ and $l'_a$ for the
upper and lower components, respectively.

\begin{figure}[!h]
       \includegraphics[width=1.0\textwidth]{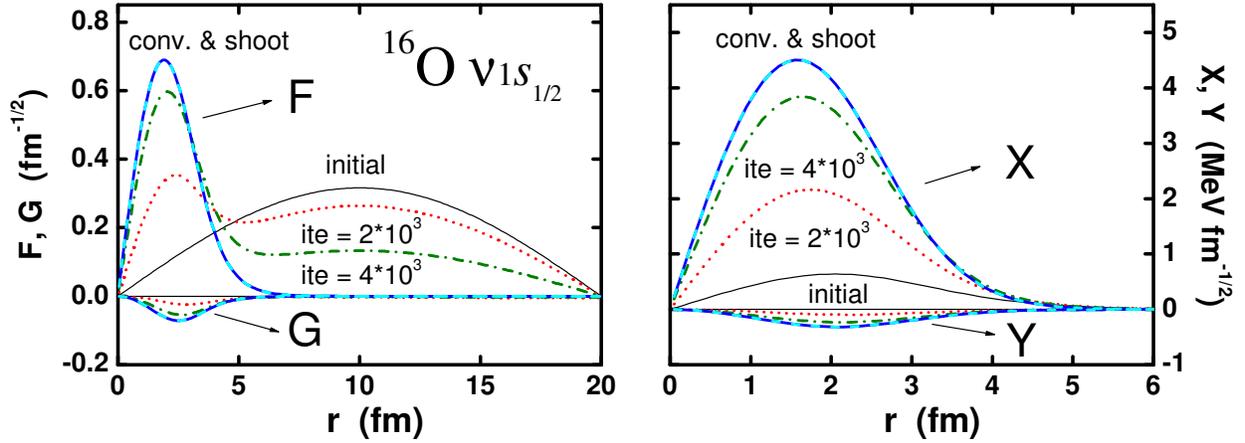} 
       \caption{ (color online)
                The evolutions of the single-particle wave
                functions $F(r)$, $G(r)$ and the nonlocal terms $X(r)$, $Y(r)$ for $\nu 1s_{1/2}$ in nucleus $^{16}$O.
                The evolutions are illustrated at the initial (thin solid lines),
                $2\times 10^3$ (dotted lines),
                $4\times 10^3$ (dash-dotted lines) iterations.
                The converged results (thick solid lines) of the ITS evolution
                are compared with the results by shooting method (dashed lines).
       }\label{Fig2}
      \end{figure}

The ITS evolutions of the single-particle wave function and the
nonlocal terms $X(r)$, $Y(r)$ for $\nu 1s_{1/2}$ in $^{16}$O are
illustrated in Fig. \ref{Fig2}. The initial single-particle wave
function which is assumed to be spherical Bessel function for the
upper component and zero for the lower component, together with the
corresponding $X(r)$ and $Y(r)$ are denoted by the thin solid line,
which are followed by the results at $2\times 10^3$ (dotted lines)
and $4\times 10^3$ (dash-dotted lines) iterations. Finally, the
evolution results converge to the same ones obtained by the shooting
method. The results from shooting method are obtained by solving the
Dirac equation with the localized effective potentials from the
converged $X(r)$ and $Y(r)$ in the ITS evolution.

\begin{figure}[!h]
       \includegraphics[width=0.6\textwidth]{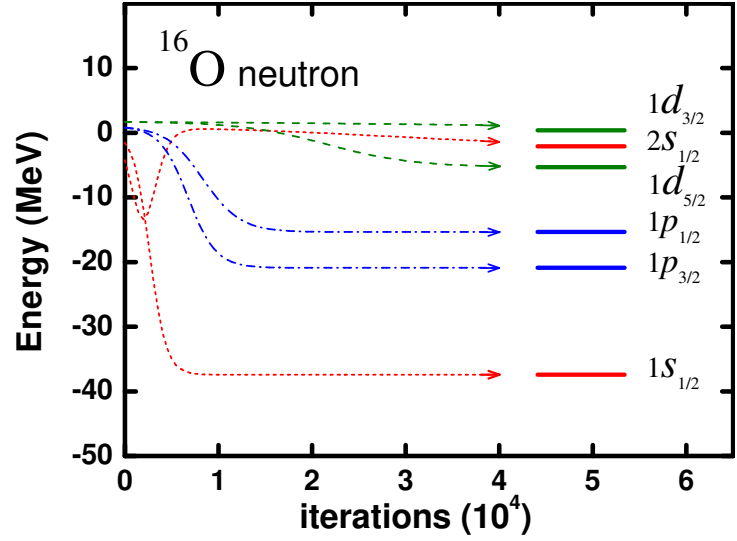} 
       \caption{ (color online)
                The evolutions of the single-particle energies and the final convergent neutron
                single-particle spectrum of nucleus $^{16}$O obtained by the ITS
                method.
       }\label{Fig3}
      \end{figure}

Similarly, the neutron single-particle spectrum of $^{16}$O can be
obtained by the ITS method as shown in Fig. \ref{Fig3}, where the
the evolutions of the corresponding single-particle energies are
also illustrated.  It can be seen that different levels converge at
different speeds. The deeper the level lies, the faster the
evolution converges.  Most levels converge monotonously except the
level $2s_{1/2}$, which evolves with fluctuation at the beginning
due to the orthonormalization with respect to the level $1s_{1/2}$.



In summary, the ITS method is applied for solving the Dirac equation
with nonlocal potential. Similar as in Ref.~\cite{ITSDirac}, the
evolution is performed for the corresponding Schr\"{o}dinger-like
equation for the upper component. It is demonstrated that the ITS
evolution can be equivalently performed for the Schr\"{o}dinger-like
equation with or without localization. The nonlocalized ITS
evolution is much more simple and efficient, and thus is recommended
for future application. These conclusions are further supported by
the numerical solution of the Dirac equation with nonlocal potential
reasonably chosen for $^{16}$O. The present algorithm provides the
same results as the shooting method for the Dirac equation with the
localized effective potentials. The investigation here demonstrates
the possibility of the ITS method in the relativistic Hartree-Fock
theory.




\end{document}